\documentclass[twocolumn,pra,floatfix,citeautoscript,nofootinbib]{revtex4-1}

\usepackage{amsbsy}
\usepackage{latexsym,epsfig,graphicx}
\usepackage{dcolumn}
\usepackage{graphicx}
\usepackage{subfigure}
\usepackage{comment}
\usepackage{color}
\usepackage{bm}
\usepackage{mathrsfs}
\usepackage{amssymb}

\usepackage[colorlinks,urlcolor=blue,citecolor=blue]{hyperref}

\begin{document}

\author{Junpeng Hou}
\author{Xi-Wang Luo}
\author{Kuei Sun}
\author{Chuanwei Zhang}
\thanks{Corresponding author. \\
Email: \href{mailto:chuanwei.zhang@utdallas.edu}{chuanwei.zhang@utdallas.edu}%
}
\affiliation{Department of Physics, The University of Texas at Dallas, Richardson, Texas
75080-3021, USA}
\title{Momentum-space Aharonov-Bohm interferometry in Rashba spin-orbit coupled
Bose-Einstein condensates}

\begin{abstract}
Since the recent experimental realization of synthetic Rashba
spin-orbit coupling paved a new avenue for exploring and
engineering topological phases in ultracold atoms, a precise,
solid detection of Berry phase has been desired for unequivocal
characterization of system topology. Here, we propose a scheme to
conduct momentum-space Aharonov-Bohm interferometry in a Rashba
spin-orbit coupled Bose-Einstein condensate with a sudden change
of in-plane Zeeman field, capable of measuring the Berry phase of
Rashba energy bands. We find that the Berry phase with the
presence of a Dirac point is directly revealed by a robust dark
interference fringe, and that as a function of external Zeeman
field is characterized by the contrast of fringes. We also build a
variational model describing the interference process with
semiclassical equations of motion of essential dynamical
quantities, which lead to agreeable trajectories and geometric
phases with the real-time simulation of Gross-Pitaevskii equation.
Our study would provide timely guidance for the experimental
detection of Berry phase in ultracold atomic systems and help
further investigation on their interference dynamics in momentum
space.
\end{abstract}

\maketitle

\section{Introduction}\label{Sec:Introduction}
Topological orders of matter have recently gained great attentions
in solid-state and cold-atom physics \cite{Duca2014,Tarruell2012,
Soltan2011,Aidelsburger2011,Cheuk2012,Alex2017,Celi2014,Cooper2011,
Lim2008,Osterloh2005,Fidkowski2013,Chen2015,Maciejko2013,Levin2006},
for their characterization of quantum phases in a different
scenario from the conventional Ginzburg-Landau orders
\cite{Hohenberg2015,Read1989} and potential application on
fault-tolerant quantum computation
\cite{Kitaev2003,Nayak2008,Sau2010}. A large variety of
topological phases, including quantum Hall states
\cite{Xiao2010,Zhang2005,Thouless1982}, topological insulators
\cite{Qi2011,Hasan2010}, and anomalous Hall states
\cite{Nagaosa2010,Haldane2004,Jungwirth2002}, can be characterized
by a geometric phase~\cite{Berry1984}, or Berry phase, of the
underlying band structure of the system. In this context, direct
measurement of Berry phase is essential for exploring the new
physics of topological states of matter.

Following the definition of Berry phase, i.e., the adiabatic phase
shift of wavefunction along a closed loop in parameter space, one
could borrow an analogous idea for the detection of Berry phase
from the Aharonov-Bohm interferometry \cite{Aharonov1959}, in
which the interference between two charged particles encircling a
magnetic flux measures the associated geometric phase in real
space. Differently, the interferometry for detecting Berry phase
on an energy band should be performed in momentum space. Although
it is a big challenge to realize the electronic interferometry in
solid-state systems, an Aharonov-Bohm interferometer in reciprocal
(lattice-momentum) space has recently been realized with ultracold
atoms in optical lattices and has successfully measured the Berry
phase of two-dimensional ($2$D) hexagonal lattice systems
\cite{Castro2009,Duca2014}.

Ultracold atomic gases exhibit great flexibility and
controllability in engineering exotic, on-demand single-particle
energy bands of a spatially continuous
system~\cite{Lin2011,Wang2012,Qu2013a,Li2017,Khamehchi2017,Qu13,Zhang13,Zhai2015,Sun2015,Demarco2015,Qu2015,Xu2015,Hou2017}
. In particular, a distinct type of single-particle band governed
by $2$D Rashba spin-orbit coupling
\cite{Hu2011,Ozawa2012,Gopalakrishnan2013,Anderson2013,Xu2012,Dudarev2004}
has been experimentally realized in both ultracold
Fermi~\cite{Huang2016,Meng2016} and Bose~\cite{Wu2016} gases.
Since such a band structure can exhibit a Dirac point, from which
crucial topological properties may emerge, probing its nontrivial
Berry phase is hence among the first experimental pursuits after
the initial realization in ultracold atoms.

\begin{figure}[b]
\centering
\includegraphics[width=0.48\textwidth]{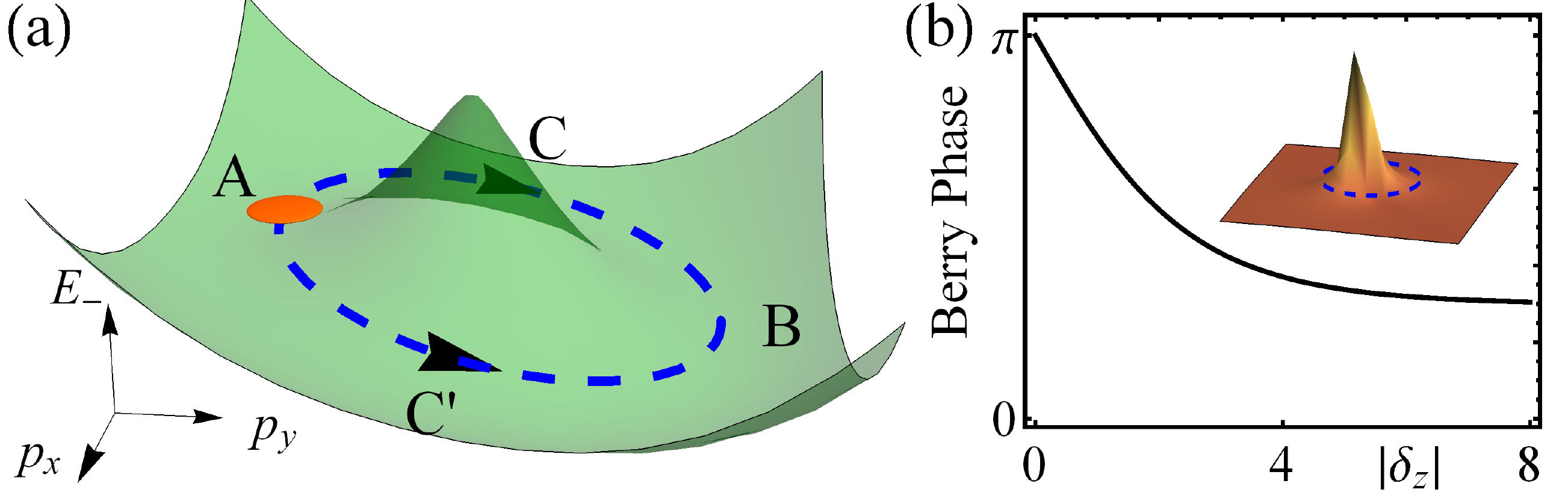}
\caption{(Color online) (a) Illustration of the momentum-space
Aharonov-Bohm interferometry. A BEC (orange spot) with Rashba
spin-orbit coupling is initially at the ground state A. After a
sudden change of detuning $\delta_x$, the deformation of energy
spectrum $E_-(p_x,p_y)$ lifts point A, and then the BEC
wavefunction packet splits into two, separately moving along paths
C and C$'$ and encountering each other at point B. The whole path
encircles nonzero Berry flux localized around the tip of the
shadow region (the Dirac point if $\delta_z=0$), resulting in a
Berry phase directly indicated by the interference pattern. (b)
Berry phase versus detuning $\delta_z$, which can be determined by
experimentally measured interference pattern through the procedure
in (a). The inset shows the schematic Berry curvature, mostly
encircled by the ring path, in the $p_x$-$p_y$ plane.}
\label{fig1}
\end{figure}

In this paper, we propose a practical scheme that conducts
momentum-space Aharonov-Bohm interferometry in Rashba spin-orbit
coupled Bose-Einstein condensates (BECs), hence measuring the
Berry phase along a loop enclosing the Dirac point. As shown in
Fig.~\ref{fig1}(a), our interferometer exploits the intrinsic ring
structure in the Rashba energy band as the interferometer loop and
the controllable detuning $\delta_x \sigma_x$ of ultracold atoms
as a trigger for driving the BEC. A sudden change of $\delta_x$
sets up the initial BEC state at the highest energy point of the
ring (point A), and as the BEC naturally pursues the lowest energy
(point B), it splits into two, following different halves of the
ring (C and C$'$) and exhibiting the interference. Since the whole
ring path encloses region with dense Berry flux, the interference
pattern would reflect the Berry phase as a function of $\delta_z$,
as in Fig.~\ref{fig1}(b). At $\delta_z=0$, the Berry phase equal
to $\pi$ indicates the presence of a Dirac point.

We adopt two complementary methods, Gross-Pitaevskii equation
(GPE) simulation and variational analysis, for studying the BEC
dynamics and interference. Our simulation shows real-time
evolution of the interference pattern for $^{87}$Rb BECs in
typical cold-atom experiments, while the variational analysis
provides an informative model capturing the key physical features
of the interference, including the trajectories in momentum and
real space as well as the geometric phase acquired during the
evolution. We also point out proper conditions for external
trapping potential and interatomic interaction under which the
interferometry procedure succeeds and discuss why improper
trapping frequency or too strong interaction sabotages the desired
dynamics for the interference. Our results ought to provide timely
guidance for ongoing experimental study on the Rashba spin-orbit
coupled quantum gases.

In Sec.~\ref{Sec:Model}, we present the model Hamiltonian and the
associated Berry phase on its lowest energy band. In
Sec.~\ref{Sec:Interferometry}, we discuss the detailed procedure
for the momentum-space Aharonov-Bohm interferometry and its
physical requirement. We then show real-time simulation results,
which reveal the relation between interference fringe contrast and
Berry phase. In Sec.~\ref{Sec:Variational}, we propose a
variational BEC wavefunction and derive its equation of motion,
which characterizes trajectories of the two splitting wave packets
as well as the accumulated geometric phase during the evolution.
Finally, we make a conclusion in Sec.~\ref{Sec:Conclusion}.


\section{Model and Hamiltonian}\label{Sec:Model}

We consider a Bose gas with atomic mass $m$ and two hyperfine spin
states ${\left( {\begin{array}{*{20}{c}} {{\psi _ \uparrow
}}&{{\psi _ \downarrow }}
\end{array}} \right)^T}$
subject to synthetic Rashba coupling in $x$-$y$ plane and tunable
Zeeman field. After integrating out the irrelevant $z$ degrees of
freedom, we write down the effective Hamiltonian,
\begin{eqnarray}
H = \frac{1}{{2m}}(\hat p_x^2 + \hat p_y^2)+ V + {H_{\rm{R}}} +
{H_{\rm{Z}}} + {H_{\rm{I}}},
\end{eqnarray}
with
\begin{eqnarray}
V &=&  \frac{m}{2}(\omega _x^2{{\hat x}^2} + \omega _y^2{{\hat y}^2}), \\
{H_{\rm{R}}} &=& {\lambda _{\rm{R}}}({{\hat p}_y}{\sigma _x} +
a{{\hat p}_x}{\sigma _y}),\label{eq:RashbaSOC} \\
{H_{\rm{Z}}} &=& {\delta _x}{\sigma _x} + {\delta _z}{\sigma _z},\\
{H_{\rm{I}}} &=& \left( {\begin{array}{*{20}{c}}
{{g_ {\uparrow\uparrow} }{{\left| {{\psi _ \uparrow }} \right|}^2} + {g_{ \uparrow  \downarrow }}{{\left| {{\psi _ \downarrow }} \right|}^2}}&0\\
0&{{g_{ \uparrow  \downarrow }}{{\left| {{\psi _ \uparrow }}
\right|}^2} + {g_ {\downarrow\downarrow} }{{\left| {{\psi _
\downarrow }} \right|}^2}}
\end{array}} \right).
\end{eqnarray}
Here $V$ is an external trapping potential, $H_{\rm{R}}$ describes
the Rashba coupling of strength $\lambda _{\rm{R}}$ and anisotropy
factor $a$, $H_{\rm{Z}}$ represents the Zeeman field $\delta_z$
($\delta_x$) in the longitudinal (transverse) direction (with the
$y$ component set to zero without loss of generality), and
$H_{\rm{I}}$ results from the spin-dependent mean-field
interaction, e.g., $ g_{\uparrow\uparrow} =
g_{\downarrow\downarrow} = 0.9554 g_{\uparrow \downarrow} \equiv
g$ in the following simulation for $^{87}$Rb systems.

Given sufficiently weak trapping, the Hamiltonian has two
single-particle energy bands in $p_x$-$p_y$ momentum space, with
the lower one being
\begin{eqnarray}
E_- = \frac{\bm{p}^2}{2m} - | {{{\vec d (\bm{p})}}} |,
\end{eqnarray}
where $\bm{p}=(p_x,p_y)$ and $\vec d = ({\lambda _{\rm{R}}}{p_x} +
{\delta _x},{\lambda _{\rm{R}}}a{p_y},{\delta _z})$. At $a=1$ and
$\delta_x=0$, the set of minima of $E_-$ form a horizontal ring
$|\bm{p}|=\sqrt{\lambda _{\rm{R}}^4-4
\delta_z^2}/(2\lambda_{\rm{R}})$ in the $p_x$-$p_y$ plane. If
$\delta_x \neq 0$, the ring structure remains but is inclined such
that it has only one maximum and one minimum at the two intercepts
with the $p_y$ axis, respectively [as points A and B in
Fig.~\ref{fig1}(a)]. Our interferometry is performed along this
Rashba ring path as we will show in Sec.~\ref{Sec:Interferometry}.

The Berry phase $\gamma$ in the region enclosed by this ring loop
can be computed as
\begin{equation}
\gamma = \oint_{L_c} \bm{A_B} \cdot d\bm{p} = \int_S
\mathcal{F}~d^2\bm{p},
\end{equation}
where $\bm{A_B} = - i \langle \zeta(\bm{p}) | \nabla_{\bm{p}} |
\zeta(\bm{p}) \rangle$ is Berry connection corresponding to
eigenstate $\zeta(\bm{p})$, $\mathcal{F} = \nabla \times \bm{A_B}$
is Berry curvature, and $L_c$ and $S$ denote the loop and the
enclosed region, respectively. For our Hamiltonian, the Berry
curvature is related to the unit vector ${\hat d}=\vec d / |\vec
d|$ as
\begin{equation}
\mathcal{F} = \frac{1}{2} \epsilon_{ij} {\hat{d}} \cdot \left(
\partial_i {\hat{d}} \times \partial_j {\hat{d}} \right),
\end{equation}
where $\epsilon_{ij}$ is the antisymmetric permutation
(Levi-Civita) symbol, and the Berry phase is equal to half the
solid angle swept by ${\hat{d}}$ in the loop integral. The Berry
phase depends on both $\delta_z$ and $\delta_x$ but is insensitive
to the latter given $\delta_x/\lambda_{\rm{R}} \ll 1$. At
$\delta_z=0$, the Berry curvature is a delta function centered at
the Dirac point $(p_x,p_y)=(0,0)$, which gives a Berry phase
$\gamma=\pi$. With $\delta_z$ increasing, the ${\hat{d}}$ vector
sweeps a less solid angle, and the Berry phase monotonically
decreases as shown in Fig.~\ref{fig1}(b). Therefore, one can
continuously change the Berry phase by tuning $\delta_z$, in
analogy of changing the magnetic flux in a conventional
Aharonov-Bohm interferometer.

\section{Interferometry and simulation}\label{Sec:Interferometry}

\begin{figure}[t]
\centering
\includegraphics[width=0.48\textwidth]{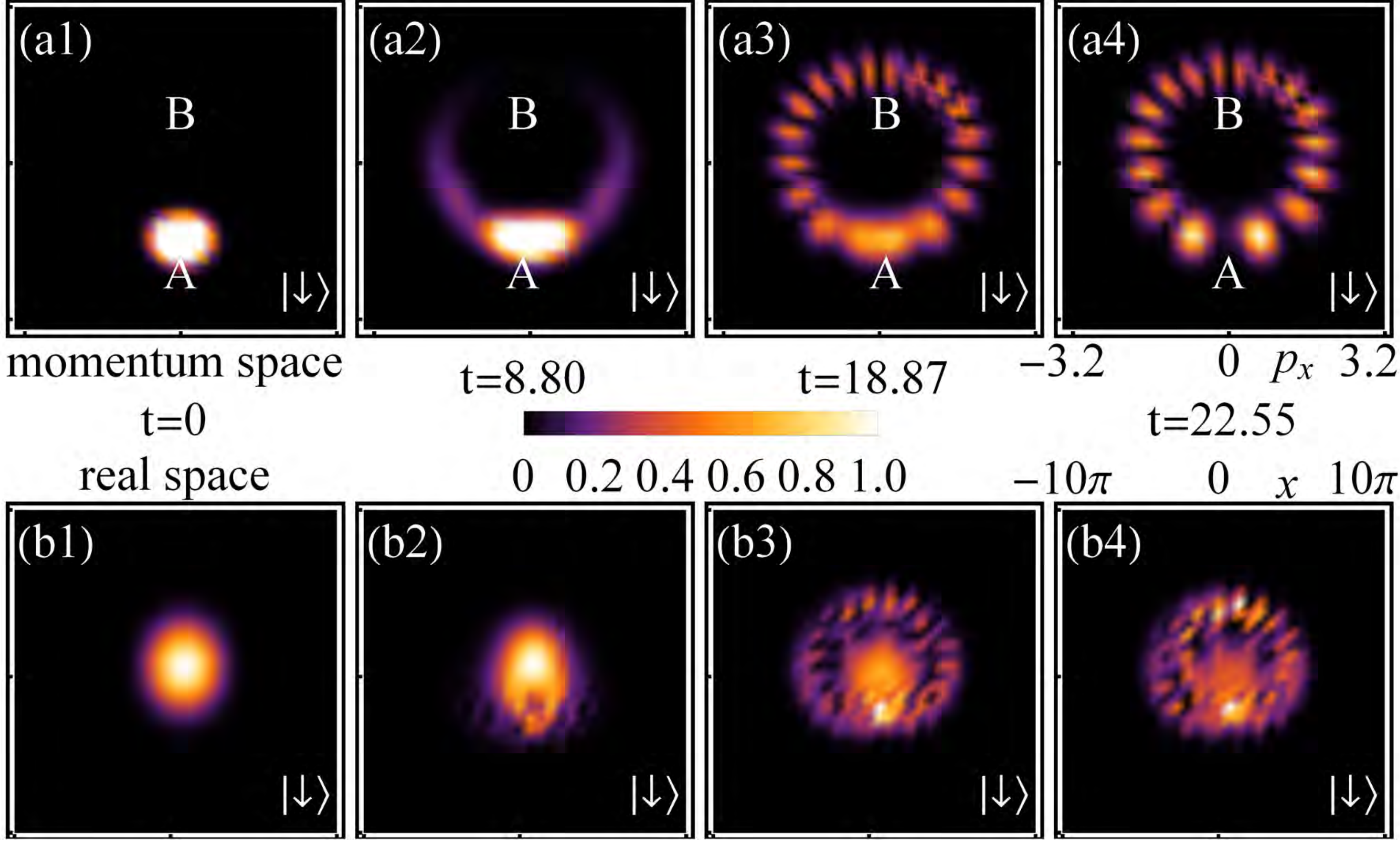}
\caption{(Color online) Time evolution (in units of ms) of
down-spin density distribution in $p_x$-$p_y$ (top) and $x$-$y$
(bottom) planes from GPE simulation (the up-spin component
exhibits same dynamics). (a1)-(a4) A ground-state BEC is first
prepared at a band minimum A. After a change of $\delta_x$ at
$t=0$, it starts to split, travel along the ring path. An
interference pattern can then be observed after the tails of
wavefunctions meet at the other side of ring (point B). (b1)-(b4)
Corresponding real-space density distributions. The parameters are
set as $\lambda _{\rm{R}}=1.5$, $a=1$, $\delta_z=1$ kHz,
$\delta_0=-350$ Hz, $\delta_1=400$ Hz, and $ng = 0.05$.}
\label{fig2}
\end{figure}

In this section, we discuss how to employ our model as an
Aharonov-Bohm interferometer for detecting the Berry phase and
present the GPE simulation results. In cold-atom systems, the
Rashba spin-orbit coupling is generated by a set of Raman lasers
that couple different spin and momentum states, and the Zeeman
energy shift is determined by the relative detunings of the lasers
\cite{Huang2016,Meng2016,Wu2016}. All the parameters are highly
tunable in the current experimental setup. We consider a trapped
BEC initially prepared at the ground state of fixed $\lambda
_{\rm{R}}$, $a$, $\delta_z$, and $\delta_x = \delta_0 < 0$, which
is a momentum Gaussian wave packet (due to the trap) centered at
the minimum of the inclined ring or $(0,-p_0)$. Then the detuning
$\delta_x$ is suddenly changed to $\delta_x = \delta_1 > 0$ (with
other parameters unchanged). The ring structure is hence inversely
inclined, such that the current location of the condensate becomes
the energy maximum [with a slight deviation $\sim O(\delta_x
\delta_z^2)$, which is negligible in our case], illustrated as
point A in Fig.~\ref{fig1}(a). As a result, the BEC wavepacket
will split into two, which follow separate paths (C and C$'$,
respectively) and move toward the new energy minimum at the
opposite end of the ring (point B). As the two waves meet and
superpose at point B, the dynamic phase cancels out due to the
symmetry between the two paths, while the geometric phase (Berry
phase), as a function of $\lambda _{\rm{R}}$, $a$, $\delta_z$, and
$\delta_x=\delta_1$, can be revealed by the density contrast of
the interference pattern.

We remark that a proper external trap is essential for driving the
motion of the condensate in momentum space since $d \hat p /dt = i
[H,\hat p]= i [V,\hat p] \neq 0$ (given the negligible
interaction). If there is no trap, any $\bm{p}$ state is a
stationary state, so the BEC does not move. However, if the
trapping potential is too strong (comparable to Rashba coupling
strength), it may also spoil the desired ground state as well as
the interference dynamics in our system \cite{Hu2012,Sinha2011}.
In addition, if the interaction is too strong, the initial BEC
wavepacket spontaneously selects one path rather than splits into
two parts. This is because the superposition of two momentum
wavepackets leads to a real-space density wave that costs too much
interaction energy. Furthermore, the Zeeman field $\delta_z$
cannot be too large. Otherwise the energy-band tip will be
flattened, and consequently, the condensate will not follow the
shallow Rashba ring groove.

Given the above constraints, our GPE simulation shows that typical
experimental parameters are indeed suited for realizing the
interference as shown in Fig.~\ref{fig2}. In panels (a1)--(a4), we
plot density distributions $\rho_{\downarrow}(\bm{p})$ in momentum
space at different time frames, with the starting point on the
ring path labeled by A and the pursued energy minimum by B.
Initially, the condensate locates at A [(a1)], which is lifted
from the ground state by the sudden change in $\delta_x$. Then the
BEC splits into two, which separately follow the ring loop [as C
and C$'$ in Fig.~\ref{fig1}(a)] toward B [(a2)]. After the two
parts encounter each other at point B, a clear ring-shape
interference pattern forms in momentum-space [(a3, a4)]. Note that
our simulation is preformed in the whole 2D plane without any
constraint in coordinates $(x,y,p_x,p_y)$. This ring structure
indeed reflects the BEC's natural motion along the ring groove
toward the lower energy in the Rashba band. Due to this petal
pattern in momentum space, the condensate also exhibits exotic
circular distributions in real-space [(b1)-(b4)]. In
Sec.~\ref{Sec:Variational}, our variational analysis of key
dynamical variables will show that the trajectories of splitting
BEC's center of mass explain the pattern exhibited by the GPE
simulation results in both momentum and real space.

\begin{figure}[t]
\centering
\includegraphics[width=0.48\textwidth]{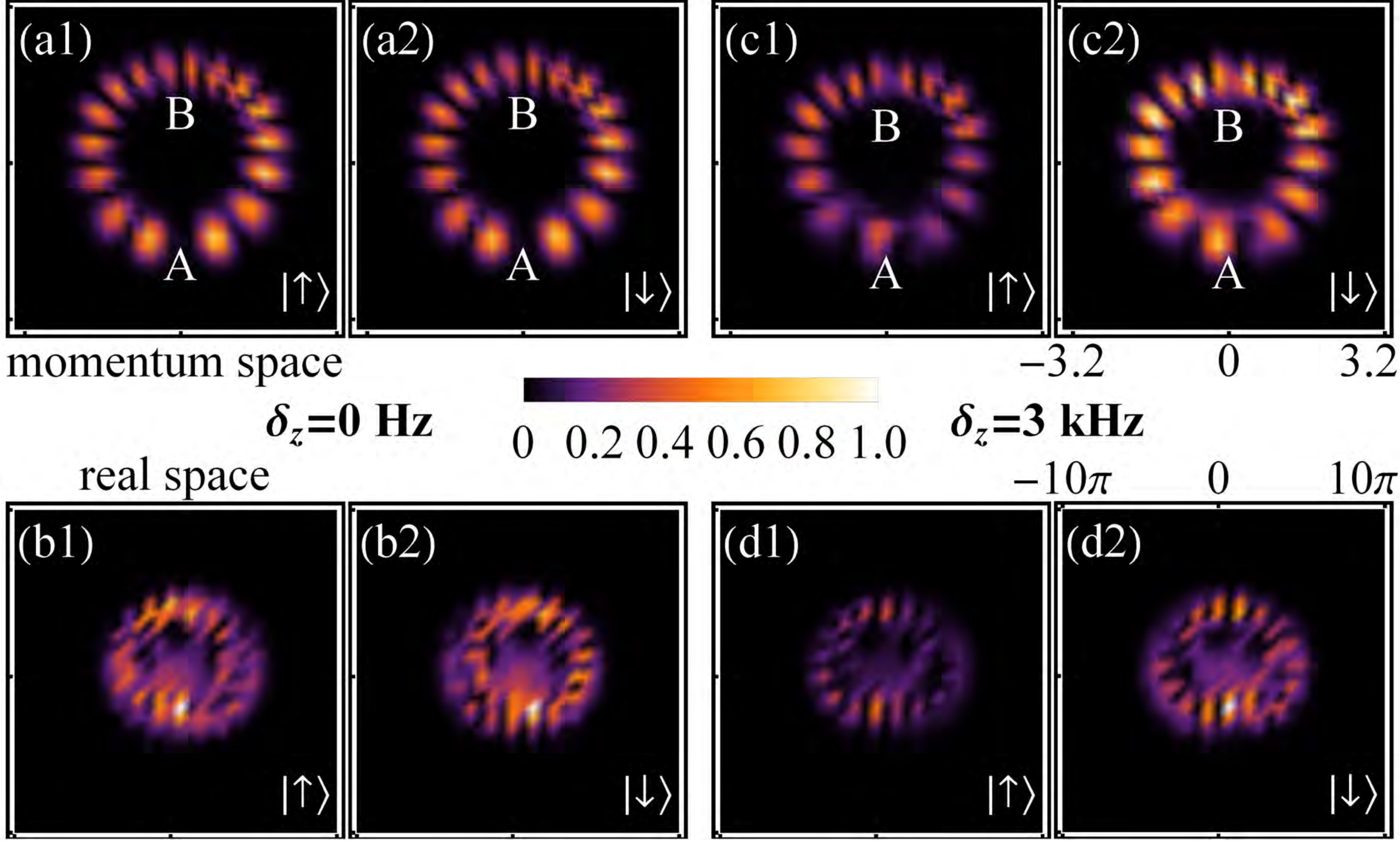}
\caption{(Color online) Momentum-space (top) and real-space (bottom)
density distributions for each spins (as labeled on the
bottom-right corner of each panel) from GPE simulation for
$\delta_z=0$ at $t=23.44$ ms (left four panels) and $\delta_z = 3$
kHz at $t=21.71$ ms (right four panels), with the other parameters
the same as in Fig.~\ref{fig2}. (a1) and (a2): at $\delta_z=0$,
there is a dark fringe (with zero density) at point B, reflecting
a $\pi$ Berry phase given by the Dirac point. (c1) and (c2): at
$\delta_z = 3$ kHz, finite density occurs at point B as the Berry
phase deviates significantly from $\pi$.} \label{fig3}
\end{figure}

Here, we turn to study the interference pattern at various Zeeman
field $\delta_z$. The GPE simulation results are presented in
Fig.~\ref{fig3}. In panels (a1) and (a2), for $\delta_z=0$, the
density at point B is constantly zero during the evolution ,
forming a robust dark fringe that indicates a $\pi$ phase shift
between the splitting BECs upon the encounter at point B. Due to
the aforementioned dynamic phase canceling, this $\pi$ phase shift
is contributed purely by the accumulated geometric phase around
the loop, thus confirming the presence of a Dirac point. Panels
(c1) and (c2) show a finite density at B for $\delta_z=3$ kHz.
This indicates the Berry phase no longer equal to $\pi$ as we
expect from the smaller solid angle swept by the $\hat d$ vector.
In Figs.~\ref{fig3} (b) and (d), we plot the corresponding
real-space density distributions. They also exhibit a roughly
ring-shape interference pattern and can be understood by
considering the group velocity and phase dynamics of the
wavepackets, as we will show in Sec.~\ref{Sec:Variational}.

\begin{figure}[t]
\centering
\includegraphics[width=0.48\textwidth]{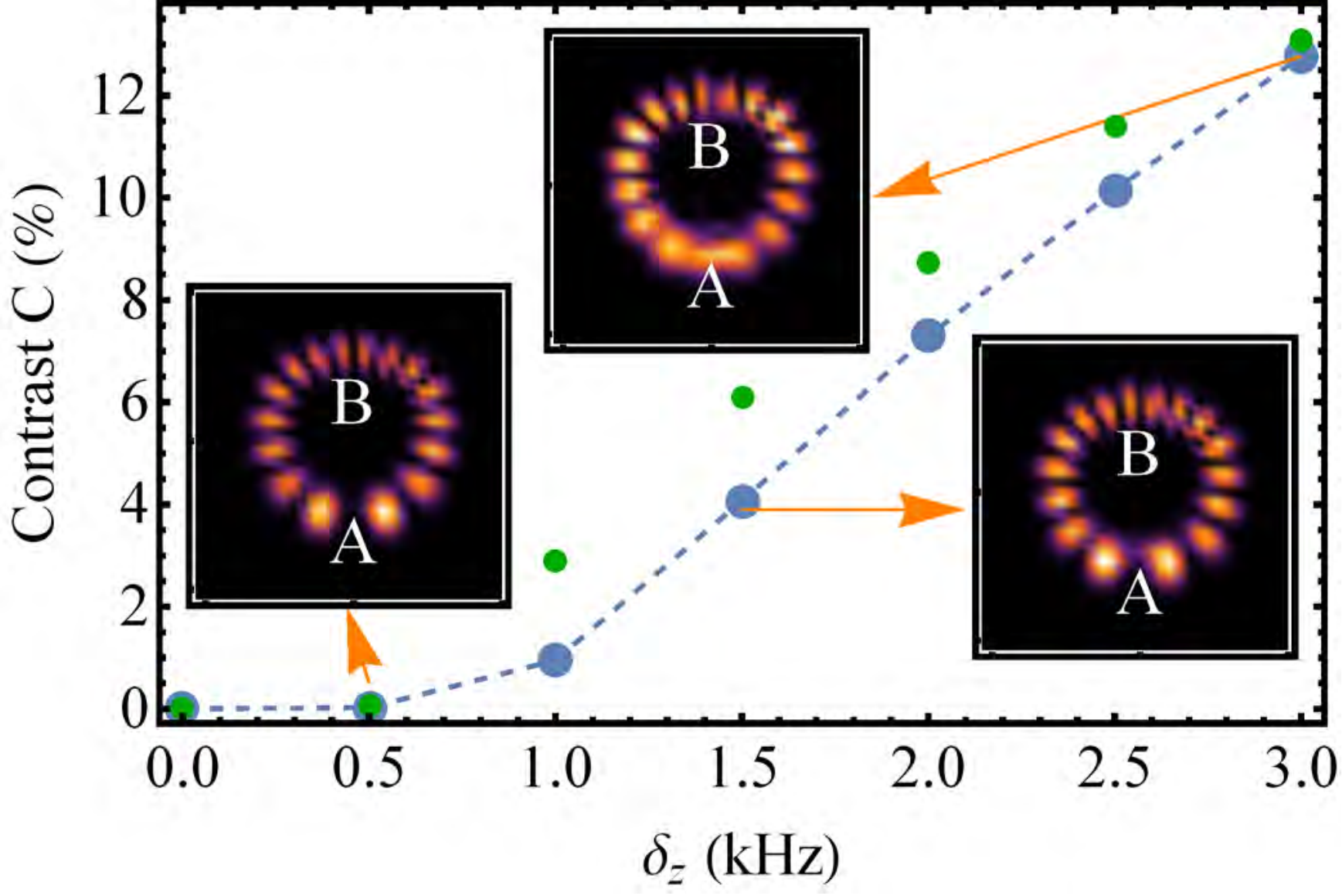}
\caption{(Color online) Fringe contrast $C$ versus $\delta_z$ from
the GPE simulation (blue) and variational analysis (green). Both
show a monotonic increase in $C$ as $\delta_z$ deviates away from
$0$. This monotonic behavior can be mapped to the monotonic trend
of Berry phase vs $\delta_z$ in Fig.~\ref{fig1}(b), such that $C$
as a measurable can be used for directly determining the Berry
phase. The insets show the momentum-space density distribution for
the corresponding data points with different contrast. All the
parameters except $\delta_z$ are the same as used in
Fig.~\ref{fig2}.} \label{fig4}
\end{figure}

In order to quantitatively relate the interference pattern with
the geometric phase, we define a local fringe contrast as
\begin{equation}
C = \frac{\int_{s_0} d\bm{p}~ \rho(\bm{p})}{\int_{s_n}
d\bm{p}~\rho(\bm{p})},\label{eq:contrast}
\end{equation}
where $\rho=\rho_\uparrow + \rho_\downarrow$ is the total density
distribution in momentum space, $s_0$ is a proper dark-fringe
region around point B, and $s_n \supset s_0$ includes the adjacent
bright-fringe regions (such that $0 \leq C \leq 1$). While $C=0$
all the time at $\delta_z=0$, it becomes nonzero as the
interference occurs for any $\delta_z>0$. Note that since the
contrast may slightly oscillates with time, in our simulation, we
record $C$ right after the center of masses of the left and right
parts of BEC both move cross the $p_x$ axis. In Fig.~\ref{fig4},
we plot $C$ vs $\delta_z$ obtained from the GPE simulation (blue
dots), which fit a monotonically increasing curve (dashed curve).
Mapped onto the monotonically decreasing relation between Berry
phase and $\delta_z$, as in Fig.~\ref{fig1}(b), this fringe
contrast $C$ may act as a good experimental measurable for
directly determining the geometric phase of the energy band.  As
we will show in Sec.~\ref{Sec:Variational}, a variational analysis
for the interference dynamics also yields a comparable
$C$-$\delta$ relation (green dots).

\begin{figure}[t]
\centering
\includegraphics[width=0.48\textwidth]{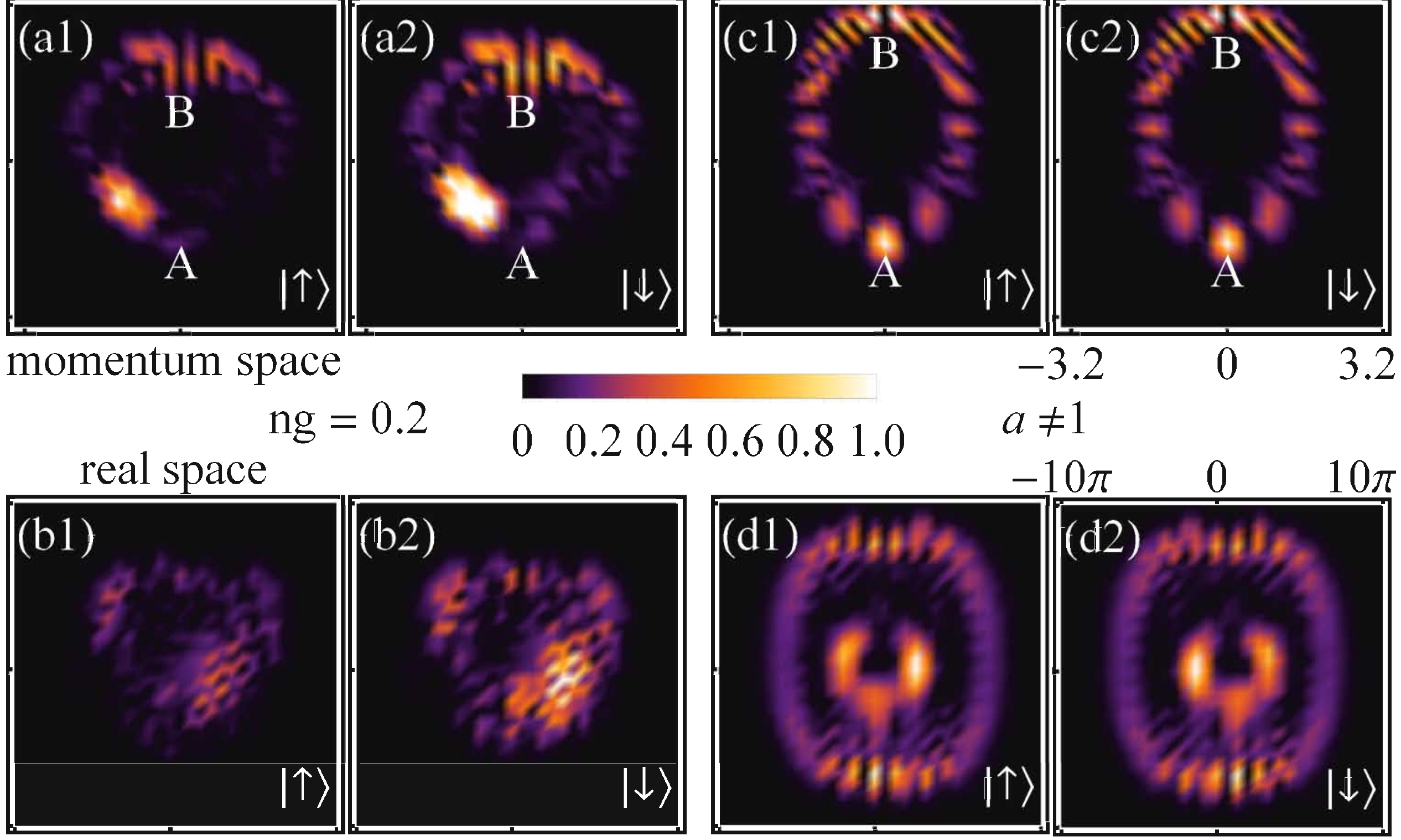}
\caption{(Color online) Momentum and real-space density
distributions for large interaction $ng=0.2$ (left four panels)
and anisotropic Rashba coupling $a=0.8$ (right four panels) from
the GPE simulation. Convention is the same as Fig.~\ref{fig3}.
(a1) and (a2): the whole condensate spontaneously choose one side
of the ring rather than split due to the interaction (at $15.41$ ms). (c1) and
(c2): a robust dark fringe still occurs at point B at anisotropic
Rashba coupling, indicating that the $\pi$ Berry phase is
independent of the deformation in energy band and interference
loop (at $32.32$ ms). The parameter changes also affect the real-space density
distribution in the bottom row. Our variational analysis confirms
the same physics.} \label{fig5}
\end{figure}

Finally, we study the effects of strong interaction and
anisotropic Rashba coupling in the experiment. As mentioned above,
a reasonably small interaction strength is favored for our scheme.
If the interaction is too strong, splitting of BEC wavefunction in
momentum space will induce real space density modulation that
highly increase the interaction energy. Consequently, the
condensate will spontaneously select one route rather than equally
split. This is shown by the GPE simulation results in
Fig.~\ref{fig5} (a1) and (a2), in which all the parameters are the
same as those in Fig.~\ref{fig3}(c,d) except $ng=0.2$ is
quadrupled. We see that the axial symmetry is broken in both
momentum and real space, and the interference fringes become more
obscure. In experiments, the Rashba coupling can be
anisotropically tuned, i.e., $a\neq 1$ in
Eq.~(\ref{eq:RashbaSOC}). As a result, the ring structure of the
energy band becomes elliptical, so does the interference loop. In
the right four panels of Fig.~\ref{fig5}, we show the GPE
simulation results for $a=0.8$, with the other parameters same as
Fig.~\ref{fig3}(a,b). We still see a dark interference fringe with
zero density at point B, indicating the $\pi$ Berry phase robust
against the deformation of energy band and interference loop, as
expected for the property of Dirac point.

\section{Variational Analysis}\label{Sec:Variational}

In this section, we reveal salient physical features of the
dynamical interference process with a simply structured
variational wave function. Since we have observed in the GPE
simulation that the BEC wavefunction intends to split into two
under weak interatomic interactions, it
is natural to consider a superposition that can describe the
splitting condensate, as
\begin{equation}
\Psi_{\rm{var}} = \phi_L e^{i\frac{\varphi}{2}}\cos\alpha+\phi_R
e^{-i\frac{\varphi}{2}}\sin\alpha, \label{eq:variational}
\end{equation}
which consists of two Gaussian wavepackets $\phi_{j=R,L}$ in the
region of $p_x>0$ and $p_x<0$, respectively. Each Gaussian
wavepacket takes a general
form~\cite{Chen2012,Zoller1996} as
\begin{eqnarray}
\phi_{j} = \zeta(\mathbf{p}_j) \prod_{\eta} && \Big[ \big(
\frac{2}{\pi R_\eta^2} \big)^\frac{1}{4} e^{-
(\frac{1}{R_\eta^2}-\frac{i}{2}\xi_\eta)(r_\eta-A_{j,\eta})^2}
\nonumber\\ && \times e^{ip_{j,\eta}(r_\eta - A_{j,\eta})}\Big],
\end{eqnarray}
where $\eta = x,y$ stands for the spatial coordinates,  $A_\eta$
is the center of mass position (in real space), $R_\eta$ is the
width of the wavepacket, and $\xi_\eta$ is introduced as the
conjugate variable for $R_\eta$, which is essential for the
completeness of this variational method~\cite{Zoller1996}. The
axial symmetry of the dynamics allows us to assume that the two
wavepackets have the same $R_\eta$ and $\xi_\eta$ (independent of
$j=L,R$), which have also been confirmed by our GPE simulation
given reasonably weak interaction. In a semiclassical picture, the
system Lagrangian $\mathscr{L} = \int d^2\bm{r} \Psi^\dagger (i
\frac{\partial}{\partial_t} - H) \Psi$ derives the equations of
motions (see details in Appendix \ref{Sec:Appendix}) as
\begin{eqnarray} \label{eom1}
\frac{d}{dt}A_{j,\eta} = \frac{\partial}{\partial
p_{j,\eta}}E_{j,-},~~\frac{d}{dt}
p_{j,\eta}=-\omega_\eta^2A_{j,\eta}, \\ \label{eom2}
\frac{d}{dt}\xi_\eta=\frac{4}{R_\eta^4}-\omega_\eta^2-\xi_\eta^2,~~\frac{d}{dt}
R_\eta=R_\eta\xi_\eta,
\end{eqnarray}
with $\alpha = \alpha_0$ being time-independent and
\begin{eqnarray} \label{eom3}
\frac{d\varphi}{dt} &=& \frac{1}{2}\Big( \frac{\partial
E_{L,-}}{\partial\alpha}\cot\alpha - \frac{\partial
E_{R,-}}{\partial\alpha}\tan\alpha \Big) \nonumber \\ &&+
\sum_{j,\eta} \epsilon_j \Big( p_{j,\eta}\frac{dA_{j,\eta}}{dt}
-\frac{1}{2}\omega_\eta^2 A_{j,\eta}^2 - E_{j,-}   \nonumber \\
&&- i \langle \zeta(\mathbf{p}) | \partial_{\mathbf{p_\eta}}
\zeta(\mathbf{p}) \rangle \frac{dp_\eta}{dt}
|_{{p_\eta}={p_{j,\eta}}} \Big),
\end{eqnarray}
where $E_{j,-}=E_-(\bm{p_j})$ and $\epsilon_{j=L,R}= \pm 1$.

\begin{figure}[t]
\centering
\includegraphics[width=0.48\textwidth]{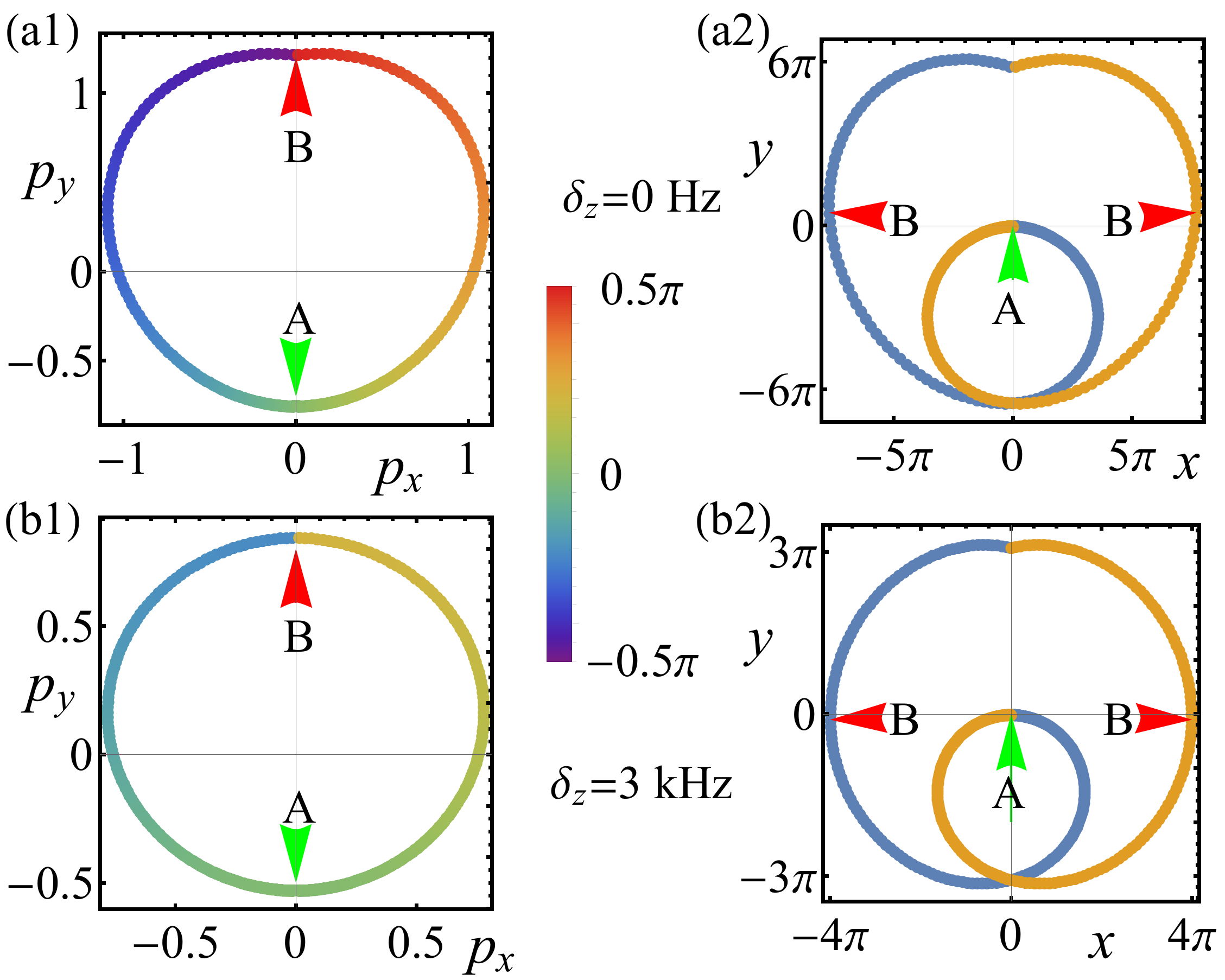}
\caption{(Color online) Center-of-mass trajectories of the two
variational wavepackets in Eq.~(\ref{eq:variational}) in momentum
space (left) and real space (right). In momentum space, the two
trajectories accumulate a relative phase (represented by colors in
bar graph) from $0$ at point A to $\pi$ at point B. In real space,
the two trajectories (blue and orange curves, respectively) form a
double-circle structure, with points A and B indicating the
corresponding positions in momentum space. (a1,a2) [(b1,b2)] are
for $\delta_z=0$ Hz ($3$ kHz). The trajectories are comparable to
the GPE simulation results in Fig.~\ref{fig3}.} \label{fig6}
\end{figure}

We choose the initial condition (at $t=0$) of Eq.~(\ref{eom1}) as
$A_{x}=A_{y}=0$ for both wavepackets (same starting point) and
small $p_{L,x} = -p_{R,x}$ for the initial velocity under slight axisymmetric
perturbation in momentum space. Such symmetry is actually
preserved by the equations of motion. The initial condition for
Eq.~(\ref{eom2}) is obtained from the minimization of system
energy functional. The trajectories generated by those equations
are presented in Fig.~\ref{fig6}, in good agreement with the GPE
results in Fig.~\ref{fig3}. Note that we also assume the equal
splitting of condensate, or $\alpha = \pi/4$, given sufficiently
weak interaction.

We turn to discuss the dynamics of phase $\varphi$ of the
variational wave function. There are three different contributions
to the time derivative of $\varphi$---the first comes from energy
terms $E_{j,-}$, the second is related to the dynamic parameters
like center-of-mass position and momentum, and the last is just
the Berry connection $\bm{A_B}$, determined together by the spinor
wavefunction and time derivative of momentum. Note that with the
axial symmetry to the $y$ (and $p_y$) axis, all the terms
related to energy and other dynamical parameters, such as $A_j^2$
and $p_j$, vanish, leaving only the last term on the RHS of
Eq.~(\ref{eom3}), which becomes
\begin{eqnarray}
\frac{d}{dt} \varphi = - i \sum_{j,\eta} \epsilon_j \left( \langle \zeta(\mathbf{p}) |
\partial_{\mathbf{p_\eta}} \zeta(\mathbf{p}) \rangle \frac{dp_\eta}{dt} |_{{p_\eta}={p_{j,\eta}}}
\right).
\end{eqnarray}
This is exactly the Berry phase defined on the ring loop since
\begin{equation}
\oint_{L_c} \bm{A_B} \cdot d\bm{p} = \oint_{L+R} \bm{A_B} \cdot d\bm{p}
= - \oint_{L} \bm{A_B} \cdot d\bm{p} +  \oint_{R} \bm{A_B} \cdot d\bm{p},
\end{equation}
where the ring-shape loop $L_c$ is divided into two parts $L_j$,
as for each part of the splitting condensate. The integral
direction of left hand side (L) is clockwise, hence carrying a
minus sign. The numerical solutions of $\varphi$ are illustrated
in Fig.~\ref{fig6} (a1) and (b1) as the curve color. In panel
(a1), the two wavepackets acquires an opposite geometric phase as
they encircle the loop. When they encounter each other at point B,
the accumulated phases are $\pm\frac{\pi}{2}$, respectively,
making a $\pi$ phase difference, which results in a dark fringe .
However, with an extra Zeeman field $\delta_z$, such phase
difference never reaches $\pi$ at point B [(a2)]. The contrast $C$
defined in Eq.~(\ref{eq:contrast}) is also evaluated by the
variation analysis, and the results (green dots) are compared with
those from GPE (blue dots) in Fig.~\ref{fig4}. They indeed show
the same monotonic trend.

We remark that the variational method well captures the physical
features of the interference dynamics as well as the geometric
phase with much fewer variables than the GPE simulation. Solving
the semiclassical equations of motion is also computationally
efficient compared with the GPE simulation. Such an analysis is
particularly useful for first searching a wide parameter region
for exotic physics, then followed by further confirmation with the
GPE simulation.

\section{Conclusion}\label{Sec:Conclusion}

We have proposed and investigated a realistic approach for
conducting momentum-space Aharonov-Bohm interferometry in Rashba
spin-orbit coupled Bose gases and shown that the interference
pattern measures the Berry phase of Rashba energy band. Our
approach utilizes the ring structure of Rashba spectrum as the
interferometry loop and the ultracold atoms tunability for
triggering the motion of BEC wavepackets along the loop. With the
real-time GPE simulation for realistic $^{87}$Rb gases, we have
found that the density contrast of the interference fringes
directly indicates the Berry phase as a monotonic function of
Zeeman detuning. In particular, the $\pi$ Berry phase of a Dirac
point (without the detuning) is exhibited by a robust dark fringe
(or zero contrast) at the end of interferometry loop.
Additionally, we have modeled the interference dynamics with a
variational wavefunction of splitting wavepackets and derived
semiclassical equations of motion for the most relevant dynamical
factors. The variational results have confirmed the trajectories
in both momentum and real space as well as the local geometric
phase acquired by the condensate along the momentum trajectory.
The complementary variational analysis and GPE simulation well
agree with each other.

Our study would provide guidance for ongoing experimental effort
measuring the Berry phase in ultracold atoms with synthetic Rashba
spin-orbit coupling. The simulated density pattern in momentum and
real space (Figs.~\ref{fig2}, \ref{fig3}, and \ref{fig5}) can be
directly compared with the time-of-flight and direct-imaging
measurements, respectively, in the experiment of $^{87}$Rb gases.
Our analysis can be extended to different experimental conditions
such as an anisotropic ring loop or a loop with several local
minima. The interferometry approach may find wide applications on
various nontrivial energy bands as well as on
high-spin~\cite{Campbell2016,Luo2015,Sun2016,Martone2016,Yu2016,Luo2017,Hu2017}
systems.

\textbf{Acknowledgements.} This work is supported by AFOSR
(FA9550-16-1-0387), NSF (PHY-1505496), and ARO (W911NF-17-1-0128).

\onecolumngrid
\appendix

\section{Semiclassical Equation of motion}\label{Sec:Appendix}
We start from a single wavepacket and study its dynamics through
deriving the semiclassical equation of motion. We assume that the
Zeeman field is relatively weak and the barrier at the center of
the Rashba ring is not flattened so that the condensate follows
the band minimum in lower band.  The normalized wavefunction
ansatz can be written as
\begin{equation}
\phi = \zeta(\mathbf{p}) \prod_{\eta=x,y} \left( \frac{2}{\pi
R_\eta^2} \right)^\frac{1}{4} e^{-
(\frac{1}{R_\eta^2}-\frac{i}{2}\xi_\eta)(r_\eta-A_\eta)^2}
e^{ip_\eta(r_\eta - A_\eta)}
\end{equation}
where $A_\eta$ is the center-of-mass position, $R_\eta$ is the
width of the wavepacket, $\xi_\eta$ is the conjugate variable for
$R_\eta$ and $\zeta(\mathbf{p})$ is normalized spin wave function
taken as the lowest eigenstate of $H_0$. Now, we can compute the
energy functional
\begin{equation}
E = E_- + \sum_{\eta=x,y} \left( \frac{1}{2R_\eta^2} + \frac{1}{8}
\xi_\eta^2R_\eta^2 + \frac{1}{2}\omega_\eta^2 \left( A_\eta^2
+\frac{1}{4}R_\eta^2 \right)\right).
\end{equation}
Note that we have ignored the interaction term at this momentum
since it plays no important role in the dynamics (we also require
this to be a weakly interacting system) and $E_-$ is the
eigenvalue of single particle Hamiltonian $H_0$. The system
Lagrangian is defined as $\mathscr{L} =\int d^2\bm{r}
\phi^*(i\frac{\partial} {\partial t} - H)\phi$, that is
\begin{equation}
\mathscr{L} =\sum_\eta \left( p_\eta\frac{d}{dt}A_\eta -
\frac{R_\eta^2}{8} \frac{d}{dt} \xi_\eta + i \langle
\zeta(\mathbf{p}) | \partial_{p_\eta} \zeta(\mathbf{p}) \rangle
\frac{d}{dt}p_\eta \right) - E,
\end{equation}
and correspondingly, EoMs are given by Lagrangian equations
\begin{eqnarray}
\frac{d}{dt}A_\eta = \frac{\partial}{\partial
p_\eta}E_-,~~\frac{d}{dt} p_\eta=-\omega_\eta^2A_\eta; \\
\nonumber
\frac{d}{dt}\xi_\eta=\frac{4}{R_\eta^4}-\omega_\eta^2-\xi_\eta^2,~~\frac{d}{dt}
R_\eta=R_\eta\xi_\eta.
\end{eqnarray}
Notice that, the term $i \langle \zeta(\mathbf{p}) |
\partial_{p_\eta} \zeta(\mathbf{p}) \rangle \frac{d}{dt}p_\eta$
does not have any contribution to the dynamics in our case,
indicating it may relate to Berry phase, which is already
demonstrated in the main text. Meanwhile, EoMs for the conjugate
pair $(A_\eta, p_\eta)$ are the same if one takes $E_-$ as the
Hamiltonian and applies Hamiltonian formalisms directly.

So far, the EoMs we got only concerns about a single dipole
motion, and thus, it is suitable to describe the oscillation or
other motion when BEC does not split into two parts (or more).

To study the Berry phase, we must have two individual wavepacket
interferences with each other. We presume that the condensate will
split into two parts $\phi_L$ (left) and $\phi_R$ (right). As they
continue to move on the ring, they will overlap at some point. In
this case, the wavefunction of the whole condensate can be written
as
\begin{equation}
\Psi_c = \Psi_L + \Psi_R= \phi_L
e^{i\frac{\varphi}{2}}\cos\alpha+\phi_R
e^{-i\frac{\varphi}{2}}\sin\alpha,
\end{equation}
where we use subscript $c$ to denote a physical quantity when it
combines both side (L and R) together and $\phi_j, j = \text{L},
\text{R}$ is defined as
\begin{equation}
\phi_j = \zeta(\mathbf{p_j}) \prod_{\eta} \left( \frac{2}{\pi
R_\eta^2} \right)^\frac{1}{4} e^{-
(\frac{1}{R_\eta^2}-\frac{i}{2}\xi_\eta)(r_\eta-A_{j,\eta})^2}
e^{ip_{j,\eta}(r_\eta - A_{j,\eta})},
\end{equation}
as the two wavepackets suppose to share the same conjugate
variables $(R_\eta, \xi_\eta)$. Typically, we have $\omega_\eta
\ll \Omega$ so that the translation symmetry is still keeping and
$\mathbf{R} \cdot (\mathbf{p_L} - \mathbf{p_R}) \gg 1$. With this
approximation, the overall energy functional is just the summation
of that of two wavepackets
\begin{equation}
E_c = E_{L,-}\cos^2\alpha + E_{R,-}\sin^2\alpha + \sum_{\eta=x,y}
\left( \frac{1}{2R_\eta^2} + \frac{1}{8} \xi_\eta^2R_\eta^2 +
\frac{1}{8}\omega_\eta^2R_\eta^2 + \frac{1}{2}\omega_\eta^2 \left(
A_{L,\eta}^2 \cos^2\alpha + A_{R,\eta}^2 \sin^2\alpha \right)
\right),
\end{equation}
where we denote $E_{j,-} = E_-(\bm{p_j})$ and this further gives
the Lagrangian
\begin{eqnarray}
\mathscr{L} _c&=&\sum_\eta \left( p_{L,\eta}
\frac{dA_{L,\eta}}{dt}  \cos^2\alpha +p_{R,\eta}
\frac{dA_{R,\eta}}{dt}  \sin^2\alpha - \frac{R_\eta^2}{8}
\frac{d}{dt} \xi_\eta \right) - \frac{1}{2}\frac{d\varphi}{dt}
\cos 2\alpha - E_c \\\nonumber &+& i \sum_\eta \left( \langle
\zeta(\mathbf{p}) | \partial_{\mathbf{p_\eta}} \zeta(\mathbf{p})
\rangle \frac{dp_\eta}{dt} |_{{p_\eta}={p_{L,\eta}}} \cos^2\alpha
+ \langle \zeta(\mathbf{p}) | \partial_{\mathbf{p_\eta}}
\zeta(\mathbf{p}) \rangle \frac{dp_\eta}{dt}
|_{{p_\eta}={p_{R,\eta}}} \sin^2\alpha \right).
\end{eqnarray}
Before proceeding to derive the equations of motion, notice that
$\mathscr{L} _c$ is independent of $\varphi$ and only has one term
$-\frac{1}{2}\frac{d\varphi}{dt}\cos 2\alpha$ containing the time
derivative of $\varphi$, indicating $\alpha$ is a constant. This
can simplify the EoMs significantly. Applying Lagrangian
equations, one may find that the dynamics of classical momentum
and position (center of mass) for left and right side are
independent of each other. Moreover, $R_\eta$ and $\xi_\eta$ still
follow the same equation as before. So, the essential part is the
relative phase $\varphi$, which is govern by
\begin{eqnarray}
\frac{d\varphi}{dt} &=& \frac{1}{2}\left( \frac{\partial
E_{L,-}}{\partial\alpha}\cot\alpha - \frac{\partial
E_{R,-}}{\partial\alpha}\tan\alpha\right) - \left(E_{L,-} -
E_{R,-}\right) \\ \nonumber &+& (p_{L,\eta}\frac{dA_{L,\eta}}{dt}
- p_{R,\eta}\frac{dA_{R,\eta}}{dt})
-\frac{1}{2}\omega_\eta^2\left( A_{L,\eta}^2 - A_{R,\eta}^2
\right) \\ \nonumber &-& i \sum_\eta \left( \langle
\zeta(\mathbf{p}) | \partial_{\mathbf{p_\eta}} \zeta(\mathbf{p})
\rangle \frac{dp_\eta}{dt} |_{{p_\eta}={p_{L,\eta}}} - \langle
\zeta(\mathbf{p}) | \partial_{\mathbf{p_\eta}} \zeta(\mathbf{p})
\rangle \frac{dp_\eta}{dt} |_{{p_\eta}={p_{R,\eta}}} \right).
\end{eqnarray}

Accounting for the interaction, one simply adds one more term in
the energy functional,
\begin{equation}
f_g = \frac{1}{\sqrt{\pi} R_x R_y} \int d^2\bm{r}~\left( g_0
|\Psi_c|^4 + g_2 |\Psi_c^* \bm{S} \Psi_c|^2 \right).
\end{equation}
However, the interaction does not affect the Berry phase here,
though it may cause a non-zero dynamic phase. Meanwhile, to avoid
interaction-induced spontaneous symmetry break, which forces the
condensate pick up one side, we require the system to have a small
interaction strength.

\twocolumngrid

\end{document}